\documentclass[aps,amsmath]{revtex4}
\usepackage{graphicx}
\begin{document}

\title{Analytic evaluation of the amplitudes for orthopositronium decay to three photons to
one-loop order}

\author{Gregory S. Adkins}
\email{gadkins@fandm.edu}
\affiliation{Franklin \& Marshall College, Lancaster, Pennsylvania 17604}
\date{\today}

\begin{abstract}
The matrix element for the decay of orthopositronium to three photons can be expressed in terms of three independent amplitudes.  We describe the analytic evaluation of these amplitudes, both to lowest order and with the inclusion of all one-loop corrections.  We use these amplitudes to find precise values for the one-loop correction to the orthopositronium decay rate $\Gamma_1=-10.286606(10) (\alpha/\pi) \Gamma_{\rm LO}$, and for the order-$\alpha^2$ ``square'' correction to the decay rate $\Gamma_2({\rm square}) = 28.860(2) (\alpha / \pi)^2 \Gamma_{\rm LO}$, where $\Gamma_{\rm LO}$ is the lowest order
rate.  We give in explicit form the function describing the one-loop correction to the distribution in phase space of the final state photons.
\end{abstract}

\pacs{12.20.Ds, 36.10.Dr}

\maketitle

\section{Introduction}

Positronium, the electron-positron bound state, is well-suited for probing many fundamental aspects of particle physics. \cite{Karshenboim04}  The physics of positronium is governed almost exclusively by the electromagnetic force---weak interaction effects are negligible compared to present experimental and theoretical uncertainties \cite{Bernreuther81,Alcorta94,Govaerts96,Czarnecki99}.  As a consequence, positronium is an ideal system for testing QED through high precision comparison between experimental and theoretical results for energy levels and decay rates.  The states of positronium are eigenstates of the charge conjugation and parity operators C and P, so positronium can be used to test the discrete symmetries C, P, and T and combinations thereof. \cite{Vetter04}  Positronium has been the focus of many past and ongoing attempts to observe physics beyond the standard model. \cite{Gninenko02,Rubbia04}  In this work we focus on the decay of spin-1 orthopositronium to three photons.

The orthopositronium decay rate has been the subject of continuing experimental and theoretical work since the first measurement by Deutsch in 1951. \cite{Deutsch51}  A summary of all experimental and theoretical results has been given by Adkins, Fell, and Sapirstein \cite{Adkins02} and updated with commentary by Rubbia \cite{Rubbia04} and by Sillou \cite{Sillou04}.  By 1990 it was apparent that there was an ``orthopositronium lifetime puzzle'', as the most precise experimental determinations (gas \cite{Westbrook89} and vacuum \cite{Nico90} results from the Michigan group) were in disagreement with theory \cite{Caswell77,Caswell79} by several standard deviations.  Many experiments were mounted to look for exotic decays of orthopositronium in an attempt to resolve the discrepancy. \cite{Rubbia04,Dobroliubov93,Dvoeglazov93,Skalsey97,Vetter04a}
Newer, somewhat less precise powder results from the Tokyo group in 1995 \cite{Asai95} and 2000 \cite{Jinnouchi00} were consistent with theory and inconsistent with the earlier Michigan results.  In 2000 the calculation of all $O(\alpha^2)$ corrections to the decay rate were completed. \cite{Adkins00,Adkins02}  Including yet higher order logarithmic corrections as well, \cite{Hill00,Kniehl00, Melnikov00} the theoretical prediction is \cite{Adkins02}
\begin{equation}
\Gamma({\rm theory}) = 7.039979(11) \mu s^{-1} \, .
\end{equation}
The $O(\alpha^2)$ correction was found to be not unusually large, leaving the discrepancy with the Michigan results intact.  Finally, in 2003 the lifetime puzzle was resolved by two new high-precision results from the Tokyo \cite{Jinnouchi03} and Michigan \cite{Vallery03} groups:
\begin{subequations}
\begin{eqnarray}
\Gamma({\rm Tokyo}) &=& 7.0396(12 {\rm \, stat.})(11 {\rm \, syst.}) \mu s^{-1} \\
\Gamma({\rm Michigan}) &=& 7.0404(10{\rm \, stat.})(8 {\rm \, syst.}) \mu s^{-1} \, ,
\end{eqnarray}
\end{subequations}
consistent with each other and with theory.

The resolution of the o-Ps lifetime puzzle does not decrease the long-term usefulness of positronium decay as a probe of fundamental physics.  Ongoing and proposed experiments involving positronium decay include those of Refs.~\cite{Vallery03,Crivelli04,Rubbia04,Gninenko04,Vetter04}.  One challenge is to improve the experimental precision of the o-Ps decay rate (currently about 200 ppm) to a level closer to the present theoretical value (about 2 ppm).  The $O(\alpha^2)$ contribution to that rate is 250 ppm, so improved experimental precision will be required in order to test the $O(\alpha^2)$ calculated result.

In this work we describe an analytic evaluation of the one-loop o-Ps$\rightarrow 3 \gamma$ decay amplitudes.  We use these amplitudes to obtain a precise value for the $O(\alpha)$ decay rate contribution, and also to calculate the part of the $O(\alpha^2)$ correction coming from the square of the one-loop amplitudes.  These results have been reported already \cite{Adkins96}---here we give further details.  We also supply an explicit analytic expression for the $O(\alpha)$ differential decay rate in terms of photon energy variables.  From the differential decay rate it is easy to obtain the $O(\alpha)$ corrected one-photon energy spectrum.  (This energy spectrum, calculated more laboriously by numerical methods, has been useful in developing simulations of experimental arrangements. \cite{Asai95,Jinnouchi03})

We adapt the formalism of covariant decay amplitudes, originally developed for the study of Z boson decay to three photons, \cite{Glover93} to the case of o-Ps$\rightarrow 3 \gamma$.  In Sec.~\ref{sec2} we use the extensive symmetries of the decay tensor to show that there are only three independent amplitudes for the o-Ps$\rightarrow 3 \gamma$ decay.  In Sec.~\ref{sec3} we express the decay amplitudes in terms of helicity variables since the spin sums are most convenient in this form.  In Sec.~\ref{sec4} the integral for the decay rate is reduced to its minimal two-dimensional form.  In Sec.~\ref{sec5} the preceeding formalism is applied to the lowest-order decay process and the lowest-order decay rate of Ore and Powell \cite{Ore49} is reproduced.  In Sec.~\ref{sec6} the method of Passarino and Veltman \cite{Passarino79} for evaluating one-loop integrals is developed.  In Sec.~\ref{sec7} the one-loop calculation is described.  Finally, in Sec.~\ref{sec8} our results for the $O(\alpha)$ and part of the $O(\alpha^2)$ decay rates are given.  The Appendix contains our explicit form for the one-loop decay distribution.

\section{Symmetries of the decay tensor}
\label{sec2}

The decay of the massive vector particle orthopositronium to three photons is described by 
the matrix element \cite{conventions}
\begin{equation}
M = \epsilon^{*}_{1 \mu_1} \, \epsilon^{*}_{2 \mu_2} \, \epsilon^{*}_{3 \mu_3}
\, \epsilon_\alpha \, M^{\mu_1 \mu_2 \mu_3 \alpha} (k_1,k_2,k_3) \quad ,
\end{equation}
where the three photons have momenta $k_i$ and polarizations $\epsilon_i$, and
the positronium atom has momentum $P = k_1+k_2+k_3$ and polarization $\epsilon$.  The decay
tensor is a linear combination of terms like $k^{\mu_1}_a k^{\mu_2}_b k^{\mu_3}_c
k^{\alpha}_d$, $k^{\mu_1}_a k^{\mu_2}_b g^{\mu_3 \alpha}$, and $g^{\mu_1 \mu_2} g^{\mu_3
\alpha}$.  The most general such tensor has 81 terms of the first type, 54 of the second,
and 3 of the third.  However, gauge invariance and Bose symmetry reduce the number of
independent contributions to only three \cite{Glover93}.  We review the argument below.

Because the decay tensor is always contracted with physical polarization vectors of
on-shell photons, which satisfy $\epsilon_{a \mu} k^\mu_a = 0$ (for $a=1$, 2, or 3), we can drop 
terms containing factors of $k^{\mu_1}_1$, $k^{\mu_2}_2$, and $k^{\mu_3}_3$.  This leaves 
only 24 terms of the first type, 30 of the second, and still 3 of the third.

By Bose symmetry, the tensor $M$ is totally symmetric under photon interchange.  This
means, for the interchange of photons 1 and 2, that
\begin{equation}
M^{\mu_1 \mu_2 \mu_3 \alpha} (k_1,k_2,k_3) = M^{\mu_2 \mu_1 \mu_3 \alpha} (k_2,k_1,k_3)
\quad.
\end{equation}
This symmetry leaves only four independent terms of the first type, six of the second,
and one of the third.  The decay tensor can be written in the manifestly symmetric way
\begin{equation}
M^{\mu_1 \mu_2 \mu_3 \alpha} (k_1,k_2,k_3) = \sum_{S_3} {\cal M}^{\mu_1 \mu_2 \mu_3 \alpha}
(k_1,k_2,k_3) \quad ,
\end{equation}
where the sum is over the six photon permutations, and the tensor $\cal M$ has the form
\begin{eqnarray}
{\cal M}^{\mu_1 \mu_2 \mu_3 \alpha} (k_1,k_2,k_3)
&=& a_1(k_1,k_2,k_3) k^{\mu_1}_3 k^{\mu_2}_1 k^{\mu_3}_1 k^{\alpha}_1
 +  a_2(k_1,k_2,k_3) k^{\mu_1}_3 k^{\mu_2}_3 k^{\mu_3}_1 k^{\alpha}_1 \cr
&+& a_3(k_1,k_2,k_3) k^{\mu_1}_3 k^{\mu_2}_3 k^{\mu_3}_2 k^{\alpha}_1
 +  a_4(k_1,k_2,k_3) k^{\mu_1}_3 k^{\mu_2}_1 k^{\mu_3}_2 k^{\alpha}_1 \cr
&+& b_1(k_1,k_2,k_3) k^{\mu_2}_1 k^{\alpha}_1 g^{\mu_1 \mu_3}
 +  b_2(k_1,k_2,k_3) k^{\mu_2}_3 k^{\alpha}_1 g^{\mu_1 \mu_3} \cr
&+& b_3(k_1,k_2,k_3) k^{\mu_1}_3 k^{\alpha}_1 g^{\mu_2 \mu_3}
 +  b_4(k_1,k_2,k_3) k^{\mu_2}_1 k^{\mu_3}_2 g^{\mu_1 \alpha} \cr
&+& b_5(k_1,k_2,k_3) k^{\mu_2}_1 k^{\mu_3}_1 g^{\mu_1 \alpha}
 +  b_6(k_1,k_2,k_3) k^{\mu_2}_3 k^{\mu_3}_2 g^{\mu_1 \alpha} \cr
&+& c(k_1,k_2,k_3) g^{\mu_1 \alpha} g^{\mu_2 \mu_3} \quad .
\end{eqnarray}
The quantities $a_i(k_1,k_2,k_3)$, $b_i(k_1,k_2,k_3)$, and $c(k_1,k_2,k_3)$ are scalar
functions of their arguments, and $b_5$, $b_6$, and $c$ are symmetric under the interchange
$k_2 \leftrightarrow k_3$.

Gauge invariance requires that the tensor $M$ be transverse
\begin{equation}
k_{1 \mu_1} M^{\mu_1 \mu_2 \mu_3 \alpha}(k_1,k_2,k_3) = 0 \quad ,
\label{transversality}
\end{equation}
with similar relations holding for contractions with $k_{2 \mu_2}$ and $k_{3 \mu_3}$.  The
condition of Eq.~(\ref{transversality}) provides 13 independent relations among the 19
variables $a_i(k_1,k_2,k_3)$, $a_i(k_1,k_3,k_2)$, $b_i(k_1,k_2,k_3)$, $b_i(k_1,k_3,k_2)$,
$b_5(k_1,k_2,k_3)$, $b_6(k_1,k_2,k_3)$, and $c(k_1,k_2,k_3)$ (with
$i=1,2,3,4$), which lead (using permutation symmetry) to three independent solutions.  So,
the tensor $\cal M$ can be expressed in terms of three scalar functions $A_1$, $A_2$, and
$A_3$ as
\begin{eqnarray}
{\cal M}^{\mu_1 \mu_2 \mu_3 \alpha} (k_1,k_2,k_3) &=& A_1 (k_1,k_2,k_3)  \frac{1}{k_1 \cdot k_3}
\Bigl (\frac{k_3^{\mu_1} k_1^{\mu_3}}{k_1 \cdot k_3}-g^{\mu_1 \mu_3} \Bigr )
k_1^\alpha
\Bigl ({\frac{k_3^{\mu_2}}{k_2 \cdot k_3}} - {\frac{k_1^{\mu_2}}{k_1 \cdot k_2}} \Bigr )
\nonumber \\
&+& A_2 (k_1,k_2,k_3)
\Bigl \{ {\frac{1}  {k_2 \cdot k_3}} 
\Bigl ({\frac{k_1^{\alpha} k_3^{\mu_1}}  {k_1 \cdot k_3}}-g^{\alpha \mu_1} \Bigr )
\Bigl ({\frac{k_1^{\mu_2} k_2^{\mu_3}}  {k_1 \cdot k_2}}-g^{\mu_2 \mu_3} \Bigr ) \nonumber \\
& & \quad \quad \quad \quad \quad \, \, \, \, + {\frac{1} {k_1 \cdot k_3}}
\Bigl ({\frac{k_1^{\mu_2}}  {k_1 \cdot k_2}} - {\frac{k_3^{\mu_2}} {k_2 \cdot k_3}} \Bigr )
( k_1^{\mu_3} g^{\alpha \mu_1} - k_1^{\alpha} g^{\mu_1 \mu_3} )
\Bigr \}
\nonumber \\
&+& A_3 (k_1,k_2,k_3)
{\frac{1} {k_1 \cdot k_3}} 
\Bigl ({\frac{k_1^{\alpha} k_3^{\mu_1}} {k_1 \cdot k_3}}-g^{\alpha \mu_1} \Bigr )
\Bigl ({\frac{k_3^{\mu_2} k_2^{\mu_3}} {k_2 \cdot k_3}}-g^{\mu_2 \mu_3} \Bigr ) \quad .
\end{eqnarray}

The amplitudes $A_1$, $A_2$, and $A_3$ can be identified by writing the decay tensor as
\begin{eqnarray}
\label{AIdentify}
M^{\mu_1 \mu_2 \mu_3 \alpha} (k_1,k_2,k_3) =
&-& A_1 (k_1,k_2,k_3) k_3^{\mu_1} k_1^{\mu_2} k_1^{\mu_3} k_1^{\alpha}
\bigl [ (k_1 \cdot k_3)^2 (k_1 \cdot k_2) \bigr ]^{-1} \nonumber \\
&+& A_2 (k_1,k_2,k_3) k_3^{\mu_1} k_1^{\mu_2} k_2^{\mu_3} k_1^{\alpha}
\bigl [ (k_1 \cdot k_2) (k_2 \cdot k_3) (k_3 \cdot k_1) \bigr ]^{-1} \nonumber \\
&+& A_3 (k_1,k_2,k_3) k_3^{\mu_1} k_3^{\mu_2} k_2^{\mu_3} k_1^{\alpha}
\bigl [ (k_1 \cdot k_3)^2 (k_2 \cdot k_3) \bigr ]^{-1} \nonumber \\
&+& \dots \quad .
\end{eqnarray}
One finds $A_1$, $A_2$, and $A_3$ by taking the coefficients of $k_3^{\mu_1} k_1^{\mu_2}
k_1^{\mu_3} k_1^{\alpha}$, $k_3^{\mu_1} k_1^{\mu_2} k_2^{\mu_3} k_1^{\alpha}$,
and $k_3^{\mu_1} k_3^{\mu_2} k_2^{\mu_3} k_1^{\alpha}$.

\section{Helicity Amplitudes}
\label{sec3}

The formula for the $\hbox{\rm o-Ps} \rightarrow 3 \gamma$ decay rate involves the absolute
square of the decay matrix element summed over final state spins and averaged over the 
initial state spin:
\begin{equation}
\overline{\vert M \vert ^2} = \sum_{\epsilon_1,\epsilon_2,\epsilon_3} \frac{1}{3}
\sum_{\epsilon} \vert M \vert ^2 \quad .
\end{equation}
This is a Lorentz invariant quantity, and can be calculated in any frame.  It is convenient
to calculate it in a two-photon rest frame.

Since we will use the orthopositronium center of mass frame for the decay rate integration,
it is useful to express our results in terms of invariant variables.  A convenient set is
given by the Mandelstam variables, which are defined by
\begin{equation}
s_{i j} = s_{j i} = (k_i+k_j)^2 = 2 k_i \cdot k_j \quad ,
\end{equation}
and satisfy
\begin{equation}
s_{i j} + s_{j k} + s_{k i} = M_{\rm Ps}^2 \quad ,
\end{equation}
where $M_{\rm Ps}$ here is the orthopositronium mass and $\{i,j,k\}$ is any permutation of
$\{1,2,3\}$.  Bar variables are defined by
\begin{equation}
\bar s_{i j} = M_{\rm Ps}^2-s_{i j} = s_{i k} + s_{j k} \quad .
\end{equation}
They satisfy
\begin{equation}
\bar s_{i j} + \bar s_{j k} + \bar s_{k i} = 2 M_{\rm Ps}^2 \quad .
\end{equation}
We note that each $s_{i j}$ and $\bar s_{i j}$ is non-negative.

We calculate $\overline{\vert M \vert ^2}$ in the $k_1 k_2$ rest frame.  The
photon and positronium momentum vectors in $(E, p_x, p_y, p_z)$ notation are given by
\begin{eqnarray}
k_1 &=& (k,0,0,k) \cr
k_2 &=& (k,0,0,-k) \cr
k_3 &=& (q,q \sin \theta,0,q \cos \theta) \cr
P &=& (E,q \sin \theta,0,q \cos \theta) \quad ,
\end{eqnarray}
where $E^2 = q^2+M_{\rm Ps}^2$.  The $k_1 k_2$ rest frame kinematic variables are given in terms of
invariants by
\begin{eqnarray}
k &=&\frac {\sqrt{s_{12}}}{2} \quad , \cr
q &=& {\frac{\bar s_{12}} {2 \sqrt{s_{12}}}} \quad , \cr
E &=& {\frac{M_{\rm Ps}^2 + s_{12}}  {\bar s_{12}}} \quad , \cr
\sin \theta &=& {\frac{2 \sqrt{s_{13} s_{23}}}  {\bar s_{12}}} \quad .
\end{eqnarray}
The helicity vectors for photon 1 are
\begin{eqnarray}
\hat e^+_1 &=& \frac{1}{\sqrt{2}} ( 0 , -1 , -i , 0) \quad , \cr
\hat e^-_1  &=& \frac{1}{\sqrt{2}} ( 0 , 1 , -i , 0) \quad .
\end{eqnarray}
For photon 2 we rotate these by $180^\circ$ around the $y$ axis using
\begin{equation}
R_2 = \begin{pmatrix}
-1&0&0 \\ 0&1&0 \\ 0&0&-1
\end{pmatrix}
\end{equation}
to find
\begin{equation}
\hat e^\pm_2 = R_2 \hat e^\pm_1 = \hat e^\mp_1 \quad .
\end{equation}
For photon 3 we rotate using
\begin{equation}
R_3 = \begin{pmatrix}
\cos \theta&0&\sin \theta \\ 0&1&0 \\ -\sin\theta&0&\cos \theta
\end{pmatrix}
\end{equation}
to find
\begin{eqnarray}
\hat e^+_3 &=& R_3 \hat e^+_1 = \frac{1}{ \sqrt{2}} (0,-\cos \theta,-i,\sin \theta) \quad ,
\cr
\hat e^-_3 &=& R_3 \hat e^-_1 = \frac{1}{ \sqrt{2}} (0,\cos \theta,-i,-\sin \theta) \quad .
\end{eqnarray}
The positronium spin $\pm 1$ helicity vectors are the same as those for photon 3:
\begin{equation}
\hat e^\pm_{Ps} = \hat e^\pm_3 \quad ,
\end{equation}
while the positronium spin 0 helicity vector is
\begin{equation}
\hat e^0_{Ps} = {\frac{1}{M_{\rm Ps}}} (q,E \sin \theta,0,E \cos \theta) \quad .
\end{equation}
The helicity amplitudes are defined by
\begin{equation}
M_{\lambda_1 \lambda_2 \lambda_3 \lambda} = \hat e^{\lambda_1 *}_{1 \mu_1}
\hat e^{\lambda_2 *}_{2 \mu_2} \hat e^{\lambda_3 *}_{3 \mu_3} 
e^{\lambda}_{Ps \alpha}M^{\mu_1 \mu_2 \mu_3 \alpha}(k_1,k_2,k_3)
\end{equation}
There are nine independent helicity amplitudes with $\lambda_1 = +$.  They are
\begin{subequations}
\begin{eqnarray}
M_{++++} &=& 2 \Bigl \{ {\frac{-A_1(123)+A_1(132)} {\bar s_{12}}} +
{\frac{A_2(123)-A_2(132)} {\bar s_{12}}} \cr
&\hbox{}& \quad -
{\frac{s_{23} A_3(132)} {s_{12} \bar s_{12}}} -
{\frac{A_3(312)} {s_{23}}} + (1 \leftrightarrow 2) \Bigr \} \quad , \\
M_{+++-} &=& 2 \Bigl \{ {\frac{A_1(123)-A_1(132)} {\bar s_{12}}} +
{\frac{-A_2(123)+A_2(132)} {\bar s_{12}}} \cr
&\hbox{}& \quad -
{\frac{A_3(123)} {s_{13}}} -
{\frac{s_{13} A_3(132)} {s_{12} \bar s_{12}}} + (1 \leftrightarrow 2) \Bigr \} \quad , \\
M_{++-+} &=& 2 \Bigl \{ -{\frac{A_1(132)} {\bar s_{12}}} + (1 \leftrightarrow 2) \Bigr \}
\quad , \\
M_{++--} &=& 2 \Bigl \{ {\frac{A_1(132)} {\bar s_{12}}} +
 {\frac{-A_2(123) - A_2(132)} {s_{23}}} -
{\frac{A_3(312)} {s_{23}}} + (1 \leftrightarrow 2) \Bigr \} \quad , \\
M_{+-++} &=& 2 \Bigl \{ {\frac{A_1(123)} {\bar s_{12}}} +
{\frac{-A_2(123)-A_2(132)} {\bar s_{12}}} +
{\frac{-A_2(312)-A_2(321)} {s_{12}}} \cr
&\hbox{}& \quad -
{\frac{A_3(213)} {s_{23}}} - {\frac{s_{23} A_3(231)} {s_{12} \bar s_{12}}}
 \Bigr \} \quad , \\
M_{+-+-} &=& 2 \Bigl \{ - {\frac{A_1(123)} {\bar s_{12}}} +
{\frac{s_{13} \bigl ( -A_2(123)-A_2(132) \bigr ) } {s_{23} \bar s_{12}}} -
{\frac{s_{13} A_3(231)} {s_{12} \bar s_{12}}} \Bigr \} \quad , \\
M_{+++0} &=& {\frac{2 \Delta} {M_{\rm Ps}^2}} \Bigl \{
{\frac{r_{13} \bigl ( A_1(123)-A_1(132) \bigr )} {\bar s_{12}}} -
\bar s_{12} A_1(312) +{\frac{r_{13} \bigl (-A_2(123)+A_2(132) \bigr )} {\bar s_{12}}}\cr
&\hbox{}& \quad  +\bar s_{12} A_2(312) +{\frac{s_{12} s_{23} A_3(123)} {s_{13}}} +
{\frac{(M_{\rm Ps}^2+s_{12}) s_{13} s_{23} A_3(132)} {s_{12} {\bar s_{12}}}} \cr
&\hbox{}& \quad +{\frac{s_{12} s_{13} A_3(312)} {s_{23}}} - (1 \leftrightarrow 2) \Bigr \} \quad , \\
M_{++-0} &=& {\frac{2 \Delta} {M_{\rm Ps}^2}} \Bigl \{ {\frac{r_{13} A_1(132) -
 r_{23} A_1(231)} {\bar s_{12}}} +
s_{12} \bigl ( A_2(123) + A_2(132) - A_2(213) - A_2(231) \bigr ) \cr
& \hbox{}& \quad - {\frac{s_{12} s_{13} A_3(312)} {s_{23}}} 
+  {\frac{s_{12} s_{23} A_3(321)} {s_{13}}} \Bigr \} \quad , \\
M_{+-+0} &=& {\frac{2 \Delta} {M_{\rm Ps}^2}} \Bigl \{ -{\frac{r_{13} A_1(123)} {\bar s_{12}}} -
\bar s_{12} A_1(321) \cr
&\hbox{}& \quad +
{\frac{(M_{\rm Ps}^2+s_{12}) s_{13} \bigl (A_2(123)+A_2(132) \bigr )} {\bar s_{12}}}
+s_{13} \bigl ( A_2(312) + A_2(321) \bigr ) \cr
&\hbox{}& \quad +
{\frac{s_{12} s_{13} A_3(213)} {s_{23}}} +
{\frac{(M_{\rm Ps}^2+s_{12}) s_{13} s_{23} A_3(231)} {s_{12} \bar s_{12}}} \Bigr \} \quad ,
\end{eqnarray}
\end{subequations}
where we have used the abbreviated notation $A_i(abc) = A_i(k_a, k_b, k_c)$ and the
definitions
\begin{eqnarray}
r_{ij} &=& M_{\rm Ps}^2 s_{ij} - s_{ik} s_{jk} \quad , \cr
\Delta &=& \sqrt{\frac{M_{\rm Ps}^2} {2 s_{12} s_{13} s_{23}}} \quad .
\end{eqnarray}
The other three $\lambda_1 = +$ amplitudes are related to the previous ones by
\begin{eqnarray}
M_{+--+} &=& M_{+-+-}(1 \leftrightarrow 2) \quad , \cr
M_{+---} &=& M_{+-++}(1 \leftrightarrow 2) \quad , \cr
M_{+--0} &=& M_{+-+0}(1 \leftrightarrow 2) \quad .
\end{eqnarray}
The $\lambda_1 = -$ amplitudes are given by the parity relations
\begin{eqnarray}
M_{- \lambda_2 \lambda_3 \pm} &=& M_{+ -\lambda_2 -\lambda_3 \mp} \quad , \cr
M_{- \lambda_2 \lambda_3 0} &=& -M_{+ -\lambda_2 -\lambda_3 0} \quad .
\end{eqnarray}
The squared decay matrix element can be written as
\begin{equation}
\overline{\vert M \vert ^2} = \frac{2}{3} \sum_{\lambda_2 , \lambda_3 , \lambda} \vert M_{+
\lambda_2 \lambda_3 \lambda} \vert ^2 \quad .
\end{equation}

\section{The Decay Rate Integral}
\label{sec4}

We will calculate the $\hbox{\rm o-Ps} \rightarrow 3 \gamma$ decay rate integral in the
positronium center of mass frame.  The decay rate integral is given by
\begin{equation}
\Gamma = {\frac{1}{3!}} {\frac{1} {2M_{\rm Ps}}} \int {\frac{d^3k_1} {(2 \pi)^3 2 \omega_1}}
{\frac{d^3k_2} {(2 \pi)^3 2 \omega_2}} {\frac{d^3k_3} {(2 \pi)^3 2 \omega_3}}
(2 \pi)^4 \delta(P-k_1-k_2-k_3) \overline{\vert M \vert ^2}
\end{equation}
where $w_i = \vert \vec k_i \vert$ are the photon energies.  Of the nine variables in $\vec
k_1$, $\vec k_2$, $\vec k_3$, four are determined in terms of the others by energy-momentum
conservation
\begin{eqnarray}
\omega_1 + \omega_2 + \omega_3 &=& M_{\rm Ps} \quad , \cr
\vec k_1 + \vec k_2 + \vec k_3 &=& 0 \quad .
\end{eqnarray}
Three variables describe the orientation in space of the decay plane.  The remaining two
variables describe the relative orientation of the photons in the decay plane.  We will use
the energies of two of the photons for this last pair of variables.  Each photon can have
any energy between $0$ and $W=M_{\rm Ps}/2$.  We find it convenient to introduce dimensionless
variables $x_i = \omega_i/W$ which satisfy $0 \le x_i \le 1$, $x_1+x_2+x_3 = 2$ and are
given in terms of invariants by $x_i = \bar s_{j k}/M_{\rm Ps}^2$.  In terms of the $x$'s, one has
\begin{equation}
\Gamma = {\frac{W} {768 \pi^3}} \int_0^1 d x_1 \int_{1-x_1}^1  d x_2 \,
\overline{\vert M \vert ^2} \quad .
\label{decayIntegral}
\end{equation}

\section{The Lowest Order Decay Rate}
\label{sec5}

The lowest order decay amplitude is given by
\begin{equation}
M_{\rm LO} = \sum_{S_3} {\rm tr} \bigl [ (-i e \gamma
\epsilon^*_3) {\frac{i}{\gamma(-P/2+k_3)-m}} (-i e \gamma \epsilon^*_2)
{\frac{i} {\gamma(P/2-k_1)-m}} (-i e \gamma \epsilon^*_1) \Psi \bigr ] 
\quad ,
\end{equation}
\begin{figure}
\includegraphics{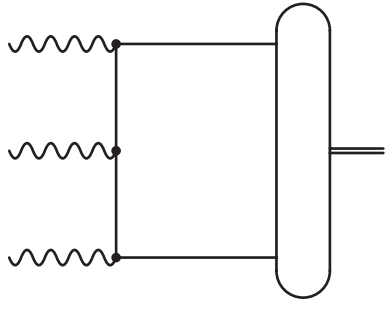}
\caption{The lowest order o-Ps $\rightarrow 3 \gamma$ decay graph.  The factor on the right represents
the initial spin-1 wave function.}
\label{fig1}
\end{figure}
where the sum is over the six permutations of the final state photons.  The wave function factor
is given by
\begin{equation}
\label{wavefuncfactor}
\Psi = \sqrt{2M_{\rm Ps}}
\begin{pmatrix} 0&\vec \sigma \cdot \hat \epsilon/\sqrt{2} \\ 0 & 0 \end{pmatrix} \phi_0 \, ,
\end{equation}
which contains the spin-1 spin factor, a normalization factor, and the wave function at
contact
\begin{equation}
\phi_0 = \sqrt{\frac{m^3 \alpha^3} {8 \pi}} \, .
\end{equation}
We write
\begin{equation}
\begin{pmatrix} 0&\vec \sigma \cdot \hat \epsilon \\ 0 & 0 \end{pmatrix} 
= \frac{1}{4} (\gamma N+1) \gamma
\epsilon (\gamma N-1)
\end{equation}
for the positronium spin factor where $N=P/(2 W)$.  The lowest order decay amplitude (see Fig.~1) becomes
\begin{equation}
M_{\rm{LO}} = {\frac{-i \pi \alpha^3} {4}} \sum_{S_3} {\frac{1} {x_1 x_3}} {\rm tr} \bigl [
\gamma
\epsilon^*_3 (-\gamma R_3+1) \gamma \epsilon^*_2 (\gamma R_1+1) \gamma \epsilon^*_1 (\gamma
N+1) \gamma \epsilon (\gamma N-1) \bigr ]
\end{equation}
where $R_i = N-K_i$, $K_i = k_i/W$ and $W \approx m$.  The lowest order decay
tensor has the corresponding form
\begin{equation}
M^{\mu_1 \mu_2 \mu_3 \alpha}_{\rm LO} = {\frac{-i \pi \alpha^3}{4}} \sum_{S_3} {\frac{1}
{x_1 x_3}} {\rm tr}
\bigl [ \gamma^{\mu_3} (-\gamma R_3+1) \gamma^{\mu_2} (\gamma R_1+1) \gamma^{\mu_1}
(\gamma N+1) \gamma^\alpha (\gamma N-1) \bigr ] \quad .
\end{equation}
We replace $N$ by $(K_1+K_2+K_3)/2$ and expand this out, and identify the $A_i$ functions
by use of Eq.~(\ref{AIdentify}).  The lowest order functions are
\begin{eqnarray}
A^{\rm LO}_1(x_1,x_2,x_3) &=& 0 \quad , \cr
A^{\rm LO}_2(x_1,x_2,x_3) &=& 16 i \pi m^2 \alpha^3 {\frac{\bar x_1 \bar x_2 \bar x_3} {x_1
x_2 x_3}} \quad , \cr
A^{\rm LO}_3(x_1,x_2,x_3) &=& 0 \quad ,
\end{eqnarray}
where $x_i =  \bar s_{j k}/M_{\rm Ps}^2$ and $\bar x_i = 1-x_i = s_{j k}/M_{\rm Ps}^2$.  
Clearly, $A^{\rm{LO}}_2$ is a factor in each helicity amplitude.  One has
\begin{eqnarray}
M^{\rm{LO}}_{++--} &=& {\frac{-x_3} {\bar x_1 \bar x_2}} \frac{A^{\rm{LO}}_2}{m^2} \quad , \cr
M^{\rm{LO}}_{+-++} &=& M^{\rm{LO}}_{+---} = {\frac{-1} {x_3 \bar x_3}} \frac{A^{\rm{LO}}_2}{m^2}
\quad , \cr
M^{\rm{LO}}_{+-+-} &=& {\frac{-\bar x_2} {\bar x_1 x_3}} \frac{A^{\rm{LO}}_2}{m^2} \quad , \cr
M^{\rm{LO}}_{+--+} &=& {\frac{- \bar x_1} {\bar x_2 x_3}} \frac{A^{\rm{LO}}_2}{m^2} \quad , \cr
M^{\rm{LO}}_{+-+0} &=& {\frac{\bar x_2} {x_3}} \sqrt{\frac{2} {\bar x_1 \bar x_2 \bar
x_3}} \frac{A^{\rm{LO}}_2}{m^2} \quad , \cr
M^{\rm{LO}}_{+--0} &=& {\frac{\bar x_1} {x_3}} \sqrt{\frac{2} {\bar x_1 \bar x_2 \bar
x_3}} \frac{A^{\rm{LO}}_2}{m^2} \quad ,
\end{eqnarray}
with all other $M^{\rm{LO}}_{+ \lambda_2 \lambda_3 \lambda}$ amplitudes equal to zero.  One
finds that
\begin{equation}
\overline{\vert M_{\rm{LO}} \vert ^2} = {\frac{512}{3}} \pi^2 \alpha^6 \Bigl \{ \Bigl (
{\frac{\bar x_1} {x_2 x_3}} \Bigr )^2 + \Bigl ( {\frac{\bar x_2} {x_1 x_3}} \Bigr )^2 +
\Bigl ( {\frac{\bar x_3} {x_1 x_2}} \Bigr )^2 \Bigr \} \quad .
\end{equation}
The lowest order decay rate is the Ore and Powell result \cite{Ore49}
\begin{eqnarray}
\Gamma_{\rm{LO}} &=& {\frac{2} {9 \pi}} m \alpha^6 \int_0^1 dx_1 \int_{1-x_1}^1 dx_2 
\Bigl \{ \Bigl ( {\frac{\bar x_1} {x_2 x_3}} \Bigr )^2 + \Bigl ( {\frac{\bar x_2} {x_1
x_3}} \Bigr )^2 + \Bigl ( {\frac{\bar x_3} {x_1 x_2}} \Bigr )^2 \Bigr \} \cr
&=& {\frac{2} {9 \pi}} (\pi^2-9) m \alpha^6 \quad .
\label{lowest_order_rate}
\end{eqnarray}

\section{One-Loop Integrals}
\label{sec6}

We used the method of Passarino and Veltman \cite{Passarino79} to evaluate the one-loop
integrals.  Since this approach is widely used, and lengthy to describe in detail, we will
just list the one-loop integrals that are required but only work the scalar integrals out in
detail.  The Passarino-Veltman method will be illustrated in the case of the three-point
functions.

The general definition of the one-loop form factors is through
\begin{eqnarray}
\bigl \{ X_0, X_\mu, X_{\mu \nu}, \cdots \bigr \} &=& \mu^{2 \epsilon} \int (dq)''_n \,
\bigl \{1, q_\mu, q_\mu q_\nu, \cdots \bigr \} \cr
&\times& \bigl [ (-q^2+m_1^2)
(-(q+p_1)^2+m_2^2) (-(q+p_1+p_2)^2+m_3^2) \cdots \bigr ]^{-1} \, .
\end{eqnarray}
Ultraviolet divergences are controlled through dimensional regularization with $n=4-2
\epsilon$ the dimensionality of spacetime.  We define $(dq)''_n = d^n q/(i \pi^{n/2})$. 
The quantity $\mu$ is a reference mass introduced with the regularization which we take
to be equal to the electron mass $m$.  Functional dependences on the masses and momenta are
indicated by $X(m_1, m_2, m_3, \cdots; p_1, p_2, \cdots)$.

The one-point function is trivially evaluated:
\begin{eqnarray}
A(m_1) &=& m^{2 \epsilon} \int (dq)''_n {\frac{1} {(-q^2+m_1^2)}} \cr
&=& -m_1^2 \Gamma(\epsilon) + \bar A(m_1) + O(\epsilon)
\end{eqnarray}
where
\begin{equation}
\bar A(m_1) = -m_1^2 \bigl [ 1-\ln(m_1^2/m^2) \bigr ] \quad .
\end{equation}

The two-point functions are defined by
\begin{equation}
\bigl \{ B_0, B_\mu, B_{\mu \nu} \bigr \} = m^{2 \epsilon} \int (dq)''_n \,
\bigl \{1, q_\mu, q_\mu q_\nu \bigr \} \bigl [ (-q^2+m_1^2) (-(q+p)^2+m_2^2) \bigr
]^{-1}
\quad .
\end{equation}
The scalar function $B_0$ is
\begin{eqnarray}
B_0(m_1, m_2; p) &=& m^{2 \epsilon} \int dx {\frac{\Gamma(\epsilon)} {\Delta^\epsilon}} \cr
&=& \Gamma(\epsilon) + \bar B_0(m_1, m_2; p)
\end{eqnarray}
where $\Delta = (1-x) m_1^2 + x m_2^2 - x(1-x) p^2$ and
\begin{equation}
\bar B_0(m_1, m_2; p) = -\int dx \ln \bigl ( \Delta / m^2 \bigr ) \quad .
\end{equation}
All parametric integrals will be taken between the limits $0$ and $1$.  The cases of
interest are
\begin{equation}
\bar B_0(0, m; p) = 2 + {\frac{1-\rho} \rho} \ln (1-\rho)
\end{equation}
where $\rho = p^2/m^2$, and
\begin{equation}
\bar B_0(m, m; p) = 2 \Bigl \{ 1-\sqrt{\frac{4-\rho} \rho} \arctan{\sqrt{\frac{\rho}{4-\rho}}}
\Bigr \} \quad ,
\end{equation}
valid for $0 \le \rho \le 4$.

The three-point functions are defined by
\begin{eqnarray}
\label{eq4A}
\bigl \{ C_0, C_\mu, C_{\mu \nu}, C_{\mu \nu \alpha} \bigr \} &=& m^{2 \epsilon} \int (dq)''_n
\, \bigl \{1, q_\mu, q_\mu q_\nu, q_\mu q_\nu q_\alpha \bigr \} \cr
&\times&  \bigl [ (-q^2+m_1^2)
(-(q+p_1)^2+m_2^2) (-(q+p_1+p_2)^2+m_3^2) \bigr ]^{-1} \, .
\end{eqnarray}
The general forms for $C_\mu$, $C_{\mu \nu}$, and $C_{\mu \nu \alpha}$ are
\begin{subequations}
\begin{eqnarray}
\label{eq4B}
C_\mu &=& p_{1 \mu} C_{1 1} + p_{2 \mu} C_{1 2} \quad , \\
C_{\mu \nu} &=& p_{1 \mu} p_{1 \nu} C_{2 1} + p_{2 \mu} p_{2 \nu} C_{2 2} + \{ p_1 p_2
\}_{\mu \nu} C_{2 3} + g_{\mu \nu} C_{2 4} \quad , \\
C_{\mu \nu \alpha} &=& p_{1 \mu} p_{1 \nu} p_{1 \alpha} C_{3 1} + p_{2 \mu} p_{2 \nu}
p_{2 \alpha} C_{3 2} + \{ p_1 p_1 p_2 \}_{\mu \nu \alpha} C_{3 3} \cr
&\hbox{}& \quad + \{ p_1 p_2 p_2 \}_{\mu
\nu \alpha} C_{3 4} + \{ p_1 g \}_{\mu \nu \alpha} C_{3 5} + \{ p_2 g \}_{\mu \nu \alpha}
C_{3 6} \quad ,
\end{eqnarray}
\end{subequations}
where
\begin{eqnarray}
\{ p k \}_{\mu \nu} &=& p_\mu k_\nu + k_\mu p_\nu \quad , \cr
\{ p p k \}_{\mu \nu \alpha} &=& p_\mu p_\nu k_\alpha + p_\mu k_\nu p_\alpha + k_\mu p_\nu
p_\alpha \quad , \cr
\{ p g \}_{\mu \nu \alpha} &=& p_\mu g_{\nu \alpha} + p_\nu g_{\mu \alpha} + p_\alpha
g_{\mu \nu} \quad .
\end{eqnarray}
The only divergent terms here are $C_{2 4}$, $C_{3 5}$, and $C_{3 6}$.

We illustrate the Passarino-Veltman procedure by describing the evaluation of $C_{1 1}$ and
$C_{1 2}$ in terms of $C_0$ and the $B$ functions.  We start by multiplying Eq.~(\ref{eq4B}) by
$p_1^\mu$ and $p_2^\mu$:
\begin{subequations}
\label{eq4C} \begin{eqnarray}
p_1^2 C_{1 1} + p_{12} C_{1 2} &=& \langle q \cdot p_1 \rangle_C \equiv R_1 \quad , \\
p_{12} C_{1 1} + p_2^2 C_{1 2} &=& \langle q \cdot p_2 \rangle_C \equiv R_2 \quad ,
\end{eqnarray}
\end{subequations}
where $p_{ij} = p_i \cdot p_j$ and $\langle \, \rangle_C$ is the integral operator on the
RHS of Eq.~(\ref{eq4A}) (so that for example $C_0 = \langle 1 \rangle_C$).  We write
Eqs.~(\ref{eq4C}) as
\begin{equation}
X \begin{pmatrix} C_{1 1} \\ C_{1 2} \end{pmatrix} = \begin{pmatrix} R_1 \\ R_2 \end{pmatrix} \quad , \quad X = \begin{pmatrix} p_1^2 & p_{12} \\ p_{12} & p_2^2 \end{pmatrix}
\end{equation}
with the solution
\begin{equation}
\begin{pmatrix} C_{1 1} \\ C_{1 2} \end{pmatrix} = X^{-1} \begin{pmatrix} R_1 \\ R_2 \end{pmatrix} \quad .
\end{equation}
We find $R_1$ and $R_2$ by noting that
\begin{subequations}
\begin{eqnarray}
q \cdot p_1 &=& {\frac{-1}{2}} \bigl \{ -(-q^2+m_1^2) + (-(q+p_1)^2+m_2^2) + f_1 \bigr \}
\quad , \\
q \cdot p_2 &=& {\frac{-1}{2}} \bigl \{ -(-(q+p_1)^2+m_2^2) + (-(q+p_1+p_2)^2+m_3^2) + f_2
\bigr \} \quad ,
\end{eqnarray}
\end{subequations}
where
\begin{subequations}
\begin{eqnarray}
f_1 &=& m_1^2 - m_2^2 + p_1^2 \quad , \\
f_2 &=& m_2^2 - m_3^2 + (p_1+p_2)^2-p_1^2 \quad .
\end{eqnarray}
\end{subequations}
Then we see that
\begin{subequations}
\begin{eqnarray}
R_1 &=& \langle q \cdot p_1 \rangle_C = {\frac{-1}{2}} \bigl \{ -B_0(m_2, m_3; p_2) +
B_0(m_1, m_3; p_1+p_2) \cr
&\hbox{}& \quad + f_1 C_0(m_1, m_2, m_3; p_1, p_2) \bigr \} \quad , \\
R_2 &=& \langle q \cdot p_2 \rangle_C = {\frac{-1}{2}} \bigl \{ -B_0(m_1, m_3; p_1+p_2)
+B_0(m_1, m_2; p_1) \cr
&\hbox{}& \quad + f_2 C_0(m_1, m_2, m_3; p_1, p_2) \bigr \} \quad .
\end{eqnarray}
\end{subequations}
Since the $B$ functions are already known, only the scalar $C_0$ function remains to be
computed.  Similarly, the $C_{2 i}$ functions can be evaluated in terms of the $C_{1 i}$'s
and $B$'s, etc.  At each level in the ladder, only the scalar functions are new.

The general three-point scalar integral is
\begin{eqnarray}
C_0(m_1,m_2,m_3;p_1,p_2) &=& \int (dq)'' \bigl [ (-q^2+m_1^2) (-(q+p_1)^2+m_2^2) \cr
&\hbox{}& \quad \quad \times (-(q+p_1+p_2)^2+m_3^2) \bigr ]^{-1} \cr
&=& \int dz dx {\frac{z}{\Delta}} \quad ,
\end{eqnarray}
where the limit $n \rightarrow 4$ has been taken since $C_0$ is ultraviolet finite, and
$(dq)'' \equiv (dq)''_4 = d^4q/(i \pi^2)$.  For $\Delta$ one finds
\begin{equation}
\Delta = (1-z) m_1^2 + z(1-x) m_2^2 + z x m_3^2 - z(1-z) p_1^2 - x z(1-z) 2 p_{12} - x
z(1-x z) p_2^2 \quad .
\end{equation}
The cases of interest here are
\begin{equation}
C_0(0,m,m;p_1,p_2) = {\frac{1}{2 p_{12}}} \Bigl \{ {\rm Li}_2 \Bigl ( {\frac{p_1^2 + 2 p_{12}}
{m^2}} \Bigr ) - {\rm Li}_2 \Bigl ( {\frac{p_1^2} {m^2}} \Bigr ) \Bigr \}
\end{equation}
which holds when $p_2^2=0$;
\begin{equation}
C_0(0,m,m;p_1,p_2) = {\frac{1} {2 m^2 \alpha}} \Bigl \{ {\rm Li}_2(1-2 \alpha) + 2
\zeta(2) - 2 \Bigl ( \arctan \sqrt{\frac{1-\alpha} \alpha} \, \, \Bigr )^2 \Bigr \} \quad ,
\end{equation}
which holds when $p_1^2 = m^2$ and $(2 p_1+p_2)^2=0$ and where $\alpha =
2+p_{12}/m^2$; and
\begin{equation}
C_0(m,m,m;p_1,p_2) = {\frac{-1} {2 p_{12}}} \Bigl \{ {\rm L} \Bigl ( {\frac{(p_1+p_2)^2}
{m^2}} \Bigr ) - {\rm L} \Bigl ( {\frac{p_1^2} {m^2}} \Bigr ) \Bigr \}
\end{equation}
where $p_2^2=0$ and
\begin{equation}
{\rm L}(s) = \int dz {\frac{\ln(1-z(1-z)s)} {(1-z)}} = -2 \Bigl ( \arctan \sqrt {\frac{s}
{4-s}}  \, \Bigr )^2 \quad .
\end{equation}
The dilogarithm function ${\rm Li}_2(x)$ is discussed in detail by Lewin. \cite{Lewin81}

The four-point functions are defined by
\begin{eqnarray}
\bigl \{ D_0, D_\mu, D_{\mu \nu}, D_{\mu \nu \alpha}, D_{\mu \nu \alpha \beta} \bigr \}
&=& \int (dq)'' \, \bigl \{1, q_\mu, q_\mu q_\nu, q_\mu q_\nu q_\alpha,
q_\mu q_\nu q_\alpha q_\beta \bigr \} \quad \quad \hbox{} \cr
&\times& \bigl [ (-q^2+m_1^2) (-(q+p_1)^2+m_2^2) (-(q+p_1+p_2)^2+m_3^2) \cr
&\times& (-(q+p_1+p_2+p_3)^2+m_4^2) \bigr ]^{-1} \quad .
\end{eqnarray}
We have dispensed with the regularization here since all of the $D$ functions needed for
our calculation are ultraviolet finite.  While general expressions for $D_0$ exist, we need
only a few special cases.  In particular, we find that
\begin{equation}
D_0(0,m,m,m;m N,-k_1,-k_2) = {\frac{8} {\sqrt{s_{12} \bar s_{12} s_{23} \bar s_{23}}}}
\Bigl \{ {\rm Li}_2 \Bigl ( {\frac{x_+} {\sqrt{D}}} , \theta \Bigr ) - 
{\rm Li}_2 \Bigl ( {\frac{-x_-} {\sqrt{D}}} , \theta \Bigr ) \Bigr \} \quad ,
\end{equation}
where
\begin{subequations}
\begin{eqnarray}
D &=& {\frac{m^2 \bar s_{12}} {s_{12} \bar s_{23}}} \quad , \\
x_{\pm} &=& {\frac{1}{2}} \Bigl ( 1 \pm \sqrt {\frac{s_{23} \bar s_{12}} {s_{12} \bar
s_{23}}}  \, \, \Bigr ) \quad , \\
\tan \theta &=& \sqrt {\frac{\bar s_{23}} {s_{23}}} \quad ,
\end{eqnarray}
\end{subequations}
and ${\rm Li}_2(r,\theta)$ is the dilogarithm of complex argument. \cite{Lewin81}
By some transformations among the momentum vectors, one can show that
\begin{equation}
D_0(0,m,m,m;m N-k_1,-k_2,-k_3) = D_0(0,m,m,m;m N,-k_3,-k_2) \quad .
\end{equation}
Finally, we also have
\begin{eqnarray}
D_0(m,m,m,m;-k_1,-k_2,-k_3) &=& {\frac{-4} {\sqrt{s_{12} \bar s_{12} s_{23} \bar s_{23}}}} 
\Bigl \{ {\rm Li}_2 \Bigl ( {\frac{y_+} {\sqrt{D_1}}},\theta_1 \Bigr ) - 
{\rm Li}_2 \Bigl ( {\frac{-y_-} {\sqrt{D_1}}},\theta_1 \Bigr ) \cr
&\hbox{}& + {\rm Li}_2 \Bigl ( {\frac{y_+} {\sqrt{D_3}}},\theta_3 \Bigr ) -
{\rm Li}_2 \Bigl ( {\frac{-y_-} {\sqrt{D_3}}},\theta_3 \Bigr ) \cr
&\hbox{}& - {\rm Li}_2 \Bigl ( {\frac{y_+} {\sqrt{D_0}}},0 \Bigr ) +
{\rm Li}_2 \Bigl ( {\frac{-y_-} {\sqrt{D_0}}},0 \Bigr ) \Bigr \} \quad ,
\end{eqnarray}
where
\begin{subequations}
\begin{eqnarray}
D_0 &=& {\frac{\bar s_{12} \bar s_{23}} {4 s_{12} s_{23}}} \quad , \\
D_1 &=& {\frac{m^2 \bar s_{23}} {s_{12} s_{23}}} \quad , \\
D_3 &=& {\frac{m^2 \bar s_{12}} {s_{12} s_{23}}} \quad , \\
y_\pm &=& {\frac{1}{2}} \bigl ( 1 \pm 2 \sqrt {D_0} \, \bigr ) \quad , \\
\theta_1 &=& \arctan \sqrt {\frac{s_{12}} {\bar s_{12}}} \quad , \\
\theta_3 &=& \arctan \sqrt {\frac{s_{23}} {\bar s_{23}}} \quad .
\end{eqnarray}
\end{subequations}
All of these integrals were done directly by way of Feynman parameters.

The five-point functions are required for the ladder diagram.  The five-point functions are
very difficult to evaluate in general.  We require only a special case, where $m_1=0$,
$m_2=m_3=m_4=m_5=m$, $p_1=m N$, $p_2=-k_1$, $p_3=-k_2$, $p_4=-k_3$.  One feature of this
special case is that there is a binding singularity: the scalar five-point function
diverges, so we will have to base our implementation of the Passarino-Veltman formalism on
the integral of the vector $q_\mu$, which is finite, instead of on the divergent scalar
integral.  Also, we have not yet evaluated the three- and four-point functions with the
necessary momenta.  We give the three-, four-, and five-point functions with the special
case mass and momenta values the names $E$, $F$, and
$G$:
\begin{subequations}
\begin{eqnarray}
\langle f \rangle_E &=& m^{2 \epsilon} \int (dq)''_n \, f \bigl [ (-q^2) (-(q+p_1)^2+m^2)
(-(q-p_1)^2+m^2) \bigr ]^{-1} \quad , \\
\langle f \rangle_{F(p_2)} &=& m^{2 \epsilon} \int (dq)''_n \, f \bigl [ (-q^2)
(-(q+p_1)^2+m^2) (-(q-p_1)^2+m^2) \cr
&\hbox{}& \quad \times (-(q+p_1+p_2)^2+m^2) \bigr ]^{-1} \quad , \\
\langle f \rangle_{G(p_2,p_3)} &=& m^{2 \epsilon} \int (dq)''_n \, f \bigr [ (-q^2)
(-(q+p_1)^2+m^2) (-(q-p_1)^2+m^2) \cr
&\hbox{}& \quad \times (-(q+p_1+p_2)^2+m^2) (-(q+p_1+p_2+p_3)^2+m^2) \bigr ]^{-1} \quad ,
\end{eqnarray}
\end{subequations}
where $p_1=p=m N$.  The first two of these are special cases of the three- and four-point
functions.  Because of the binding singularity, $E_0 = \langle 1 \rangle_E$, $F_0 = \langle
1 \rangle_F$, and $G_0 = \langle 1 \rangle_G$ all diverge.  We start our analysis with the
vector integrals $E_\mu = \langle q_\mu \rangle_E$, $F_\mu = \langle q_\mu \rangle_F$, and
$G_\mu = \langle q_\mu \rangle_G$.

The three-point special case vector integral has the general form
\begin{equation}
E_\mu = \langle q_\mu \rangle_E = p_\mu E_1 \quad .
\end{equation}
It is not hard to show (by use of symmetric integration) that $E_1=0$, so that
\begin{equation}
E_\mu = 0 \quad .
\end{equation}

The four-point special case vector integral has the general form
\begin{equation}
F_\mu(p_2) = \langle q_\mu \rangle_F = p_{1 \mu} F_{11}(p_2) + p_{2 \mu} F_{12}(p_2)
\quad .
\end{equation}
The necessary vector integrals for $p_2=-k_1$ are
\begin{subequations}
\begin{eqnarray}
F_{1 1}(-k_1) &=& {\frac{-1} {4 x_1^2}} \Bigl \{ {\rm Li}_2(1-2 x_1) - 2 x_1 \ln( 2 x_1 )
-2 \theta^2 - 4 \sqrt{x_1 \bar x_1} \theta + 2 \zeta(2) \Bigr \} \quad , \\
F_{1 2}(-k_1) &=& {\frac{1} {x_1}} F_{1 1}(-k_1) + {\frac{1} {8 x_1^2}} \bigl [ 2 {\rm
Li}_2(1-2 x_1) + \zeta(2) - 2 \theta^2 \bigr ] \quad 
\end{eqnarray}
\end{subequations}
where
\begin{equation}
\theta = \arctan{\sqrt{\frac{\bar x_1} {x_1}}} \quad .
\end{equation}
When $p_2=-k_1-k_2$ one finds
\begin{equation}
F_{\mu}(-k_1-k_2) = -F_{\mu}(-k_3) \quad ,
\end{equation}
which implies that
\begin{eqnarray}
F_{1 1}(-k_1-k_2) &=& -F_{1 1}(-k_3) + 2 F_{1  2}(-k_3) \quad , \cr
F_{1 2}(-k_2-k_2) &=& F_{1 2}(-k_3) \quad .
\end{eqnarray}

The five-point special case vector integral has the general form
\begin{equation}
G_\mu(-k_1,-k_2) = \langle q_\mu \rangle_G = p_\mu G_{11}+k_{1 \mu} G_{12}+k_{3 \mu}
G_{13} \quad .
\end{equation}
The $G_{1 i}$ functions are given by
\begin{subequations}
\begin{eqnarray}
G_{11}(x_1,x_3) &=& \frac{1}{8 \bar x_1} \bigl [ I_0(x_1,x_3)+I_1(x_1,x_3) \bigr ]
- \frac{1}{8 \bar x_3} \bigl [ I_0(x_3,x_1)+I_1(x_3,x_1) \bigr ] \quad , \\
G_{12}(x_1,x_3) &=& \frac{1}{16 x_1 \bar x_1} \bigl [(1-2 x_1) I_0(x_1,x_3) - I_1(x_1,x_3)
\bigr ] \cr
&\hbox{}& \quad + \frac{1}{16 x_1 \bar x_3} \bigl [ I_0(x_3,x_1) + I_1(x_3,x_1) \bigr ] \quad , \\
G_{13}(x_1,x_3) &=& \frac{-1}{16 \bar x_1 x_3} \bigl [I_0(x_1,x_3) + I_1(x_1,x_3)\bigr ] \cr
&\hbox{}& \quad -\frac{1}{16 x_3 \bar x_3} \bigl [ (1-2 x_3) I_0(x_3,x_1) - I_1(x_3,x_1) \bigr ] \quad ,
\end{eqnarray}
\end{subequations}
where
\begin{subequations}
\begin{eqnarray}
I_0(x_1,x_3) &=& \frac{1}{\sqrt{x_1 \bar x_1 x_3 \bar x_3}} \bigl [ {\rm Li}_2(r_+,\theta) -
{\rm Li}_2(r_-,\theta) \bigr ] \quad , \\
I_1(x_1,x_3) &=& \frac{1}{(x_1-x_3)} \ln \Bigl ( \frac{x_1}{x_3} \Bigr ) - \frac{2}{\sqrt{x_3
\bar x_3}} \arctan \Bigl ( \sqrt{\frac{\bar x_3}{x_3}} \Bigr ) \quad ,
\end{eqnarray}
\end{subequations}
with $r_\pm = \sqrt{\bar x_1} \pm \sqrt{x_1 \bar x_3 / x_3}$ and $\theta = \arctan
\sqrt{x_1 / \bar x_1}$.

\section{Analysis of the one-loop decay diagrams}
\label{sec7}

The decay amplitudes can be written as
\begin{equation}
A_i = A_i^{(0)} + A_i^{(1)} + A_i^{(2)} + \cdots
\end{equation}
for $i=1,2,3$, where the superscript indicates the power of $\alpha$ above that of the
lowest order amplitudes $A_i^{(0)} = A_i^{LO}$.  (Terms of order $A_i^{(2)}$ and higher
also contain factors of $\ln(1/\alpha)$.)  The expressions for the squares
$\vert M_{\lambda_1,\lambda_2,\lambda_3;m} \vert ^2$ contain parts of the form
\begin{equation}
A_i^* A_j = A_i^{(0)*} A_j^{(0)} + \bigl [ A_i^{(0)*} A_j^{(1)} + A_i^{(1)*} A_j^{(0)}
\bigr ] + A_i^{(1)*} A_j^{(1)} + \bigl [ A_i^{(0)*} A_j^{(2)} + A_i^{(2)*} A_j^{(0)} \bigr ]
+ \cdots
\end{equation}
for various combinations of $i$ and $j$.  The $A_i^{(0)*} A_j^{(0)}$ terms give the
lowest-order differential decay distribution.  The $A_i^{(0)*} A_j^{(1)} + A_i^{(1)*}
A_j^{(0)}$ terms give the order-$\alpha$ correction, and the $A_i^{(1)*} A_j^{(1)}$ and
$A_i^{(0)*} A_j^{(2)} + A_i^{(2)*} A_j^{(0)}$ terms give the order-$\alpha^2$ corrections.

The graphs contributing to the order-$\alpha$ corrected decay amplitudes $A^{(1)}_i$ in the renormalized Feynman gauge are shown in Fig.~2.  The infrared divergence induced by mass-shell renormalization is regulated by use of a photon mass $\lambda$.  The self-energy (Fig.~2a) and vertex graphs (Figs.~2b, 2c) contain infrared divergences of the form $\ln \lambda M_{LO}$.  The ladder graph Fig.~2e requires special care in its evaluation since it contains an infrared binding singularity.  This divergence can be identified and subtracted out, as discussed in detail in Ref. \cite{Adkins02}.  The result is that
\begin{equation}
M_L = \Bigl \{ \frac{\pi}{\lambda} + \ln \lambda -1+O(\lambda) \Bigr \} \Bigl ( \frac{\alpha}{\pi} \Bigr ) M_{LO} + M_{LS} \quad .
\label{expression_for_the_ladder}
\end{equation}
The subtracted ladder graph is
\begin{equation}
M_{LS} = -i \alpha^4 m^2 \sum_{S_3} \int (d \ell)'' \bigl [
\ell^2 (\ell^2-2 \ell p) (\ell^2+2 \ell p) Z(\ell) \bigr ]^{-1} \Bigl \{ \bigl (
{\rm tr}(\ell) - {\rm tr}(0)
\bigr )  -{\frac{{\rm tr}(0)} {Z(0)}} \bigl ( Z(\ell) - Z(0) \bigr ) \Bigr \} \quad ,
\label{subtracted_ladder}
\end{equation}
\begin{figure}
\includegraphics{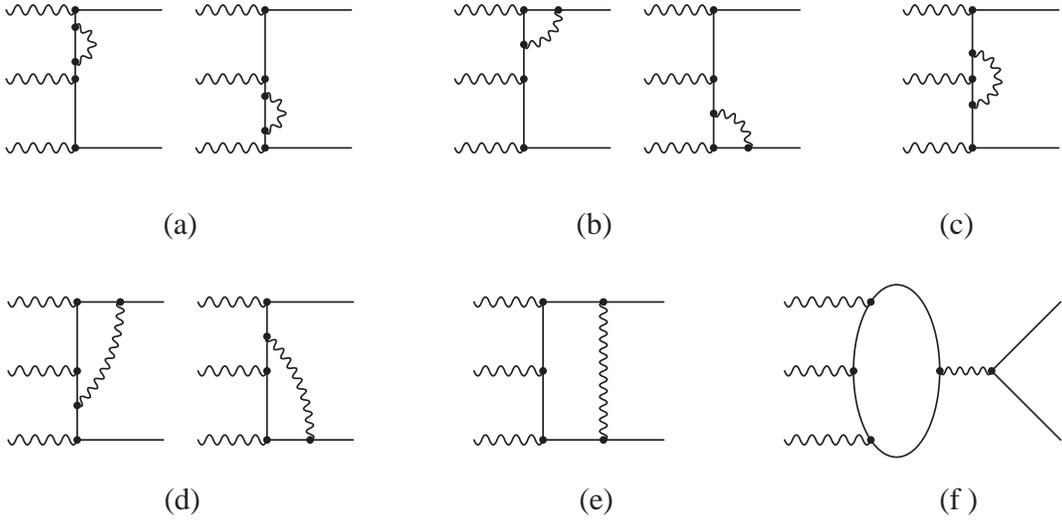}
\caption{Graphs contributing to the o-Ps decay amplitudes through order-$\alpha$.  They are
the (a) self-energy, (b) outer vertex, (c) inner vertex, (d) double vertex, (e) ladder, and
(f) annihilation contributions.  The wave function factors are implicit in these graphs.}
\label{fig2}
\end{figure}
with $p=m N$,
\begin{eqnarray}
{\rm tr}(\ell) &=& {\frac{1}{4}} {\rm tr} \bigl [ \gamma^\mu (\gamma(\ell-p)+m) \gamma
\epsilon^*_3 (\gamma(\ell-p+k_3)+m) \gamma \epsilon^*_2 (\gamma(\ell+p-k_1)+m) \cr
&\hbox{}& \times \gamma
\epsilon^*_1 (\gamma(\ell+p)+m) \gamma_\mu (\gamma N+1) \gamma \epsilon (\gamma N-1) \bigr ]
\quad ,
\end{eqnarray}
and
\begin{equation}
Z(\ell) = ((\ell-p+k_3)^2-m^2) ((\ell+p-k_1)^2-m^2) \quad .
\end{equation}
The subtraction in Eq.(\ref{subtracted_ladder}) takes away the $\ell$-independent part of
${\rm tr}(\ell)/Z(\ell)$, which would have had an infrared singularity.  This binding
singularity, regulated by the photon mass, is displayed in Eq.~(\ref{expression_for_the_ladder}).
The contributions of the order-$\alpha$ decay graphs were evaluated one by one and
summed.  The $1/\lambda$ binding singularity was removed according to the usual procedure of NRQED \cite{Caswell86,Adkins02}.  The $\ln \lambda$ terms cancel between the self-energy, vertex, and ladder graphs.  The remaining expressions are a finite sums  of rational functions of the $x_i$ times logarithms, dilogarithms, and inverse tangent functions.

\section{Results and Conclusions}
\label{sec8}

We use our analytic results for the order-$\alpha$ decay amplitudes $A^{(1)}_i$ to calculate the
order-$\alpha$ correction to the o-Ps $\rightarrow 3 \gamma$ decay rate and a part of the
order-$\alpha^2$ correction.  The individual amplitudes are quite lengthy and will not be displayed.  A simplified form for the complete order-$\alpha$ decay rate contribution is given in the Appendix.  The result for the order-$\alpha$ decay rate is \cite{integration_ref_1}
\begin{equation}
\Gamma_{1} = -10.286606(10) {\frac{\alpha}{\pi}} \Gamma_{LO} \quad .
\end{equation}
This represents a 60-fold improvement in precision over the previous best result $-10.2866(6)$ \cite{Adkins92} done using a higher dimensional integration.  The two-dimensional integral for the part of the order-$\alpha^2$ correction to the decay rate coming from the $A_i^{(1)*} A_j^{(1)}$ terms gives \cite{integration_ref_2}
\begin{equation}
\Gamma_{2}({\rm square}) = 28.860(2) \Bigl ( {\frac{\alpha}{\pi}} \Bigr )^2 \Gamma_{LO} \quad.
\end{equation}
The previous result for this contribution was $28.8(2)$. \cite{Burichenko93}

In this work we obtained analytic expressions for the o-Ps$\rightarrow 3 \gamma$ decay amplitudes.  We used these expressions to obtain precise results for the one-loop and ``square'' decay rate contributions, which were incorporated into the full calculation of two-loop corrections to the
o-Ps$\rightarrow 3 \gamma$ decay rate. \cite{Adkins00,Adkins02}  We also give an explicit form for the one-loop decay distribution (see the Appendix) which can be used to obtain the one-loop energy spectrum in a convenient form.

\begin{acknowledgments}
I am grateful for the assistance of Kunal Das in an early stage of this work, and to Zvi
Bern, Richard Fell, Russell Kauffman, Andrew Morgan, and Jonathan Sapirstein for useful
conversations.  I thank Aditya Narayanan, Sharmini Ilankovan, and Aba Mensah-Brown for helping to check formulas.  I appreciate the hospitality of the Physics Department at UCLA, where part
of this work was done, and acknowledge the support of the National Science Foundation
(through grant No. PHY-9408215) and of the Franklin and Marshall College Grants Committee.
\end{acknowledgments}

\appendix*

\section{The one-loop correction}

In the appendix we present the integral for the one-loop correction to the decay rate in
compact form.  From this integral the one-loop phase-space distribution and energy
spectrum can be obtained.  We note that for very soft photons additional effects must be taken into account in order to obtain accurate results for the phase-space distribution and energy spectrum. \cite{Pestieau02,Manohar04,Voloshin04,Ruiz04}

The one-loop correction to the decay rate is
\begin{equation}
\Gamma_1 = {\frac{m \alpha^7} {36 \pi^2}} \int_0^1 dx_1 \int_{1-x_1}^1 dx_2 \, {\frac{1}
{x_1 x_2 x_3}} \bigl \{ F(x_1,x_3) + {\rm permutations} \bigr \} \quad ,
\end{equation}
where $x_1+x_2+x_3=2$ and the ``permutations'' are the six permutations of the  variables $x_1$, $x_2$, $x_3$.  The one-loop phase space distribution is just the integrand.  The energy spectrum is found by integrating over $x_2$ but not $x_1=E_1/m$.  (The corresponding lowest-order expression is given in Eq.~(\ref{lowest_order_rate}).)  The function $F(x_1,x_3)$ is given by
\begin{equation}
F(x_1,x_3) = g_0(x_1,x_3) + \sum_{i=1}^5 g_i(x_1,x_3) h_i(x_1) + \sum_{i=6}^7 g_i(x_1,x_3)
h_i(x_1,x_3) \quad .
\end{equation}
The $h$ functions are given by
\begin{subequations}
\begin{eqnarray}
h_1(x_1) &=& \ln(2 x_1) \quad , \\
h_2(x_1) &=& \sqrt{\frac{x_1} {\bar x_1}} \theta_1 \quad , \\
h_3(x_1) &=& {\frac{1} {2 x_1}} \Bigl \{ \zeta(2)-{\rm Li}_2(1-2 x_1) \Bigr \} \quad , \\
h_4(x_1) &=& {\frac{1} {2 x_1}} \Bigl \{ \bigl ( {\frac{\pi}{2}} \bigr )^2-\theta_1^2 \Bigr \}
 \quad , \\
h_5(x_1) &=& {\frac{1} {2 \bar x_1}} \theta_1^2 \quad , \\
h_6(x_1,x_3) &=& \frac{1}{\sqrt{x_1 \bar x_1 x_3 \bar x_3}} \Bigl \{ {\rm Li}_2(r_A^+,\bar
\theta_1) - {\rm Li}_2(r_A^-,\bar \theta_1) \Bigr \} \quad , \\
\nonumber h_7(x_1,x_3) &=& \frac{1}{ \sqrt{x_1 \bar x_1 x_3 \bar x_3}} \Bigl \{ {\rm Li}_2(r_B^+,
\theta_1) - {\rm Li}_2( r_B^-, \theta_1)\\
&\hbox{}& \quad  - \frac{1}{2} {\rm
Li}_2(r_C^+,0) + \frac{1}{2} {\rm Li}_2(r_C^-,0) \Bigr \}  \quad ,
\end{eqnarray}
\end{subequations}
where $\bar x_i = 1-x_i$ and
\begin{subequations}
\begin{eqnarray}
\theta_1 &=& \arctan \bigl ( \sqrt{\bar x_1 / x_1} \, \, \bigr ) \quad , \\
\bar \theta_1 &=& \arctan \bigl ( \sqrt{x_1 / \bar x_1} \, \, \bigr ) \quad , \\
r_A^{\pm} &=& \sqrt{\bar x_1} \Bigl ( 1 \pm \sqrt{\frac{x_1 \bar x_3}  {\bar x_1 x_3}}
\, \Bigr ) \quad , \\
r_B^{\pm} &=& \sqrt{x_1} \Bigl ( 1 \pm \sqrt{\frac{\bar x_1 \bar x_3} {x_1 x_3}}
\, \Bigr ) \quad , \\
r_C^{\pm} &=& r_B^{\pm}/\sqrt{x_1} \quad .
\end{eqnarray}
\end{subequations}
The $g$ functions are given in terms of $x^{mn} = x_1^m x_3^n$ and $\bar x_2=x_1+x_3-1$ as
\begin{subequations}
\begin{eqnarray}
g_0(x_1,x_3) &=& {\frac{1} {9 x_1 \bar x_1 (1-2x_1) x_3 \bar x_3 (1-2x_3)}}
\Bigl \{ -180+2196x^{10}-4968x^{20} \cr
&\hbox{}& +5292x^{30}-2664x^{40}+504x^{50}-5848x^{11}+22639x^{21}-20280x^{31} \cr
&\hbox{}& +8405x^{41}-1240x^{51}-24x^{61}-17551x^{22}+22982x^{32}-5857x^{42} \cr
&\hbox{}& +264x^{52}+48x^{62}-3776x^{33}-878x^{43}+400x^{53}+536x^{44} \Bigr \}
\quad , \\
g_1(x_1,x_3) &=& {\frac{4}{x_1^2 (1-2 x_1)^2 (x_1-x_3) x_3}} \Bigl \{ 2 x^{20}-13
x^{30}+35x^{40}-36x^{50}+8x^{60} \cr
&\hbox{}& +4x^{70}+9x^{11}-59x^{21}+149x^{31}-210x^{41}+162x^{51}-51x^{61}+x^{71} \cr
&\hbox{}& -4x^{02}+3x^{12}+55x^{22}-126x^{32}+104x^{42}-39x^{52}+x^{62}+8x^{03} \cr
&\hbox{}& -26x^{13}+7x^{23}+22x^{33}+2x^{43}-2x^{53}-4x^{04}
+14x^{14} -8x^{24}-8x^{34} \Bigr \} \quad , \\
g_2(x_1,x_3) &=&{\frac{2} {3 x_1^3 \bar x_1 x_3}} \Bigl \{ -48x^{10}+180x^{20}
-276x^{30}+228x^{40}-108x^{50}+24x^{60} \cr
&\hbox{}& +48x^{01}-48x^{11}-144x^{31}+244x^{41}-106x^{51}+2x^{61}+4x^{71}-96x^{02} \cr
&\hbox{}& +156x^{12}-108x^{22}+168x^{32}-132x^{42}+7x^{52}+6x^{62}+48x^{03}-60x^{13} \cr
&\hbox{}& -36x^{23}+42x^{33}+9x^{43}-6x^{53}+6x^{34}-4x^{44} \Bigr \} \quad , \\
g_3(x_1,x_3) &=& {\frac{4} {x_1^2 (x_1-x_3) x_3}} \Bigl \{ -2x^{20}-2x^{40}
-4x^{60}+5x^{11}-6x^{21}+14x^{31}-4x^{41} \cr
&\hbox{}& +18x^{51}-x^{61}-4x^{02}-2x^{12}+4x^{22}-2x^{32}-26x^{42}-x^{52}+8x^{03} \cr
&\hbox{}& -7x^{13}-2x^{23}+12x^{33}+2x^{43}-4x^{04}+4x^{14} \Bigr \} \quad , \\
g_4(x_1,x_3) &=& {\frac{8}{x_1^2}} \Bigl \{ -4+7x^{10}-7x^{20}+12x^{30}
-10x^{40} +2x^{50}+8x^{01}-10x^{11} \cr
&\hbox{}& +3x^{21}-3x^{31}+2x^{41}-4x^{02}+3x^{12}+2x^{22}+x^{32} \Bigr \} \quad , \\
g_5(x_1,x_3) &=& {\frac{2 \bar x_1}  {x_1}} \Bigl \{ 8-34x^{10}+29x^{20}-4x^{30}
+6x^{11}+8x^{02}-4x^{12} \Bigr \} \quad , \\
g_6(x_1,x_3) &=& {\frac{1}{x_1 \bar x_2 x_3}} \Bigl \{ 16-76x^{10}+136x^{20}
-124x^{30}+64x^{40}-16x^{50}-60x^{01} \cr
&\hbox{}& +272x^{11}-424x^{21}+294x^{31}-104x^{41}+22x^{51}+92x^{02}-392x^{12} \cr
&\hbox{}& +484x^{22}-187x^{32}+13x^{42}+2x^{52}-76x^{03}+294x^{13}-259x^{23}+30x^{33} \cr
&\hbox{}& +3x^{43}+36x^{04}-120x^{14}+61x^{24}+3x^{34}-8x^{05}+22x^{15}
+2x^{25} \Bigr \} \quad , \\
g_7(x_1,x_3) &=& {\frac{1} {\bar x_2}} \Bigl \{ 16-48x^{10}+46x^{20}-12x^{30}
-2x^{40}-48x^{01}+60x^{11}+9x^{21} \cr
\nonumber &\hbox{}& -31x^{31}+10x^{41}+46x^{02}+9x^{12}-42x^{22}+11x^{32}-12x^{03}-31x^{13} \\
&\hbox{}& +11x^{23}-2x^{04}+10x^{14} \Bigr \} \quad .
\end{eqnarray}
\end{subequations}
\break


\begin{thebibliography}{99}
\bibitem{integration_ref_2} Quadruple precision was required in order to retain numerical
significance in some parts of the $\Gamma_2({\rm square})$ integrand.  The reported result
was obtained with approximately $10^5$ function calls.
\end{thebibliography}
\end{document}